\documentstyle[11pt,epsfig,menu97]{article}

\newcommand{\beq}{\begin{equation}}
\newcommand{\eeq}{\end{equation}}
\newcommand{\beqa}{\begin{eqnarray}}
\newcommand{\eeqa}{\end{eqnarray}}

\newcommand{\ve}{\varepsilon}

\newcommand{\krig}[1]{\stackrel{\circ}{#1}}

\begin{document}
    \setlength{\baselineskip}{2.6ex}

\title{Baryon Chiral Perturbation Theory}
\author{Ulf-G. Mei\ss ner\\
{\em FZ J\"ulich, IKP (Th), D-52425 J\"ulich, Germany}}

\maketitle

\vspace{-2.5cm}

\hfill KFA-IKP(TH)-1997-15

\vspace{1.9cm}

\begin{abstract}
\setlength{\baselineskip}{2.6ex}
\noindent I review recent progress made in the description of pion--nucleon
scattering, the $\sigma$--term, the scalar form factor and the 
strangeness content of
the nucleon in the framework of heavy baryon chiral perturbation theory.
Other topics are single pion production by real and virtual photons
as well as in proton--proton collisions.
\end{abstract}

\setlength{\baselineskip}{2.6ex}

\section*{INTRODUCTION}

\noindent The chiral Ward identities of QCD can be solved 
perturbatively at low energies by
means of an effective chiral Lagrangian, i.e. in chiral perturbation 
theory~\cite{heiri}. Formally, one considers the generating functional
of QCD,
\begin{equation}
\exp(i{\cal Z}_{\rm QCD}[v,a,s,p]) = \int [DU] \exp \biggl( i  \int
d^4 x \, {\cal L}_{\rm eff} (U;v,a,s,p) \biggr)
\,\, ,
\end{equation}
where $U$ collects the Goldstone boson fields and $v,a,s,p$ are
external fields coupled to the pertinent quark currents.
One can thus explore in a systematic fashion the
strictures of the spontaneously broken chiral symmetry of QCD. In the
presence of matter fields, like e.g. the nucleons, this amounts to a
\un{triple expansion} in powers of small momenta, quark masses and the
inverse nucleon mass at energies smaller than the chiral symmetry scale
$\Lambda_\chi \simeq 1$~GeV. The set of these small expansion parameters is
collectively denoted as $p$. One is thus deeling with a full--fledged
quantum field theory including loop graphs and renormalization. The 
theory is, however, only valid below some limiting energy and thus
renormalization to all orders is not the issue, but rather
order--by--order renormalization. At each order, new terms with
coefficients not fixed by symmetry appear, the so--called low--energy
constants (LECs). These parameters in a way parametrize our ignorance
of fully solving QCD. It is, however, important to stress that the
number of processes one looks at by far outnumbers these LECs and one
thus can make truely \un{quantitative} predictions. A great variety
of calculations
concerning processes with one, two or many nucleons have been
performed over the last few years as reviewed e.g. in~\cite{bkmr}.
Here, I will focus on some novel developments pertinent to this
meeting, in particular pion--nucleon scattering and pion production
by real and virtual photons. I will also add some remarks on the
much discussed reactions $pp \to pp\pi^0$ and $pp \to d \pi^+$.

\section*{$\pi$N--SCATTERING, $\sigma$--TERM AND ALL THAT}

\noindent A nice discussion about the relation between low--energy
pion--nucleon scattering data, the $\sigma$--term and strangeness in
the proton has been given by Gasser at the 2nd $\pi$N
symposium~\cite{juerg}. I will update here what has happened in this
field over the last few years.

\subsection*{Pion--nucleon scattering threshold parameters}
\noindent  The effective 
pion--nucleon Lagrangian consists of a string of terms with increasing 
dimension. At second order, it contains seven LECs. While these have
been determined before~\cite{bkmr}, these determinations involved some quantities 
in which large cancellations appear inducing some sizeable uncertainty.
In ref.\cite{bkma}, the four LECs related to pion--nucleon scattering and 
isoscalar--scalar external sources (as measured e.g. in the $\sigma$--term) 
were fixed from a set of nine observables which to one--loop order $p^3$
are given entirely by tree and loop diagrams with insertions from the 
dimension one and two parts of the effective Lagrangian. The fifth LEC is
only non--vanishing in case of unequal light quark masses and can thus be
estimated from the strong contribution to the neutron--proton mass difference.
The other two LECs are given by the anomalous magnetic moments of the proton
and the neutron. In that paper, it was also shown that the numerical values
of these LECs can indeed be understood from resonance exchange, however, in
some cases there is sizeable uncertainty related to certain $\Delta$ couplings.
Most interesting is the finding that the LEC $c_1$ reveals the strong pionic
correlations coupled to nucleons well known from phenomenological models of the
nucleon--nucleon force. 
The one-loop contribution to the $\pi N$-scattering amplitude 
to order $p^3$ has first been worked out by Moj\v zi\v s~\cite{moj}. Here, I 
follow ref.\cite{bkma} in which certain aspects of pion--nucleon scattering have 
also been addressed. In the center-of-mass frame the $\pi
N$-scattering amplitude $\pi^a(q) + N(p) \to \pi^b(q') + N(p')$ takes the
following form: 
\begin{equation} 
T^{ba}_{\pi N} = \delta^{ba} \Big[ g^+(\omega,t)+ i \vec
\sigma \cdot(\vec q\,'\times \vec q\,) \, h^+(\omega,t) \Big] +i \epsilon^{bac}
\tau^c \Big[ g^-(\omega,t)+ i \vec \sigma \cdot(\vec q\,'\times \vec q\,) \,
h^-(\omega,t) \Big] 
\end{equation}
with $\omega = v\cdot q = v\cdot q\,'$ the pion cms energy and $t=(q-q\,')^2$ 
the invariant momentum transfer squared. $g^\pm(\omega,t)$ refers to the
isoscalar/isovector non-spin-flip amplitude and $h^\pm(\omega,t)$ to the
isoscalar/isovector spin-flip amplitude. After renormalization of the pion
decay constant $F_\pi$ and the pion-nucleon coupling constant $g_{\pi N}$, one
can give the one-loop contributions to the cms amplitudes
$g^\pm(\omega,t)$ and $h^\pm(\omega,t)$ at order $p^3$ in closed form, 
see ref.\cite{bkma}. In table 2, I show the predictions 
for the remaining S, P, D and
F-wave threshold parameters which were not used in the fit to determine the
LECs. In some cases, contributions from the dimension three Lagrangian appear.
The corresponding LECs have been estimated using resonance exchange. In 
particular, the 10\% difference in the P--wave scattering volumina $P_1^-$
and $P_2^+$ is a clear indication of chiral loops, because
nucleon and $\Delta$ Born
terms give the same contribution to these two observables. Note also that
the eight D-- and F--wave threshold parameters to this order are free of
contributions from dimension three and thus uniquely predicted. The overall
agreement of the predictions with the existing experimental values is rather
satisfactory. As pointed out by Mojzis~\cite{moj}, the chiral
expansion is moderate for most of these threshold parameters. One
should therefore perform an order $p^4$ calculation.

\renewcommand{\arraystretch}{1.3}

\begin{table}[h]
\caption{Threshold parameters predicted by CHPT. The order
of the prediction is also given together with the experimental values.}
\vspace{0.2cm}\hspace{1.6cm}
\begin{tabular}{|c|c|c|c|c|c|c|}
    \hline  
    Obs. & CHPT &  Order & Ref. & Exp. value & Ref. & Units \\
    \hline
$a^-$ &  $9.2\pm 0.4$ & $p^4$ & [5] & $ 8.4 \ldots 10.4$ &
[6] & $10^{-2}\, M_\pi^{-1} $ \\
$b^-$ &  $2.01$ & $p^3$ & [4] & $ 1.32 \pm 0.62$ &
[7] & $10^{-2}\, M_\pi^{-1} $ \\
   \hline
$P_1^-$ &  $-2.44\pm 0.13$ & $p^3$ & [4] & $ -2.52 \pm 0.03$ &
[7] & $ M_\pi^{-3} $ \\
$P_2^+$ &  $-2.70\pm 0.12$ & $p^3$ & [4] & $ -2.74 \pm 0.03$ &
[7] & $ M_\pi^{-3} $ \\
   \hline
$a^+_{2+}$ &  $-1.83$ & $p^3$ & [4] & $ -1.8 \pm 0.3$ &
[7] & $10^{-3}\, M_\pi^{-5} $ \\
$a^+_{2-}$ &  $2.38$ & $p^3$ & [4] & $ 2.20 \pm 0.33$ &
[7] & $10^{-3}\, M_\pi^{-5} $ \\
$a^-_{2+}$ &  $3.21$ & $p^3$ & [4] & $ 3.20 \pm 0.13$ &
[7] & $10^{-3}\, M_\pi^{-5} $ \\
$a^-_{2-}$ &  $-0.21$ & $p^3$ & [4] & $ 0.10 \pm 0.15$ &
[7] & $10^{-3}\, M_\pi^{-5} $ \\
   \hline
$a^+_{3+}$ &  $0.29$ & $p^3$ & [4] & $ 0.43 $ &
[7] & $10^{-3}\, M_\pi^{-7} $ \\
$a^+_{3-}$ &  $0.06$ & $p^3$ & [4] & $ 0.15 \pm 0.12$ &
[7] & $10^{-3}\, M_\pi^{-7} $ \\
$a^-_{3+}$ &  $-0.20$ & $p^3$ & [4] & $ -0.25 \pm 0.02$ &
[7] & $10^{-3}\, M_\pi^{-7} $ \\
$a^-_{3-}$ &  $0.06$ & $p^3$ & [4] & $ 0.10 \pm 0.02$ &
[7] & $10^{-3}\, M_\pi^{-7} $ \\
   \hline   
  \end{tabular} \end{table} 


\subsection*{Baryon masses and $\sigma$--term}

\noindent The scalar sector of baryon CHPT is particularly interesting since
it is sensitive to scalar--isoscalar operators and thus directly to the
symmetry breaking of QCD. This is most obvious for the pion-- and 
kaon--nucleon $\sigma$--terms, which measure the strength of the scalar
quark condensates $\bar q q$ in the proton. 
Furthermore, the quark mass
expansion of the baryon masses allows to gives bounds on the ratios of the
light quark masses~\cite{jg}. 
Only recently the results of a  calculation including  all terms of second 
order in the light quark masses, ${\cal O}(m_q^2)$, were presented~\cite{bm}. 
The calculations were performed  in the isospin limit $m_u = m_d$ and the 
electromagnetic corrections were neglected. 
The quark mass expansion of the octet baryon masses takes the form
\begin{equation}
m = \, \, \krig{m} + \sum_q \, B_q \, m_q + \sum_q \, C_q \, m_q^{3/2} + 
\sum_q \, D_q \, m_q^2  + \ldots
\label{massform}
\end{equation}
modulo logs. Here, $\krig{m}$ is the octet mass in the chiral limit of
vanishing quark masses and the coefficients $B_q, C_q, D_q$ are 
state--dependent. Furthermore, they include contributions proportional
to some LECs which appear beyond leading order in the effective Lagrangian.
In contrast to the ${\cal O}(p^3)$ calculation, which gives the leading
non-analytic terms $\sim m_q^{3/2}$, the order $p^4$ one is no longer finite
and thus needs renormalization. Intimately connected to the baryon masses
is the $\sigma$--term,
$\sigma_{\pi N} (t)  =  \hat m \, \langle p' \, 
| \bar u u + \bar d d| \, p \rangle$,
with $|p\rangle$ a proton state with four--momentum $p$ and $t 
= (p'-p)^2$ the 
invariant momentum transfer squared. A relation between $\sigma_{\pi N} (0)$
and the nucleon mass is provided by  the Feynman--Hellmann theorem, 
$\hat{m}(\partial m_N / \partial \hat{m}) = \sigma_{\pi N} (0)$, with 
$\hat m$ the average light quark mass. Furthermore, the strangeness fraction 
$y$ and $\hat \sigma$ are defined via
$
y = \frac{ 2 \, \langle p| \bar  s s|p \rangle}
{\langle p|\bar u u + \bar d d |p \rangle} \equiv 1 -
\frac{\hat \sigma}{\sigma_{\pi N} (0)} \,\, .
$
Let me turn to the calculations presented in~\cite{bm}. As shown 
in~\cite{bm}, there
are ten LECs related to symmetry breaking. Since there do not exist enough 
data to fix all these, they were estimated  by means of resonance exchange. 
To deal with such scalar-isoscalar operators, the standard resonance 
saturation picture based on tree graphs was  extended to include loop
diagrams. In~\cite{bm}  a consistent scheme to implement 
resonance exchange under such circumstances was developed. In particular, it
avoids double--counting and abids to the strictures from analyticity. Within 
the one--loop approximation and to leading order in the resonance masses, the 
analytic pieces of the pertinent graphs are still divergent, i.e.
one is left with three a priori undetermined renormalization constants
($\beta_\Delta$, $\delta_\Delta$ and $\beta_R$). These have to be
determined together with the finite scalar meson--baryon couplings $F_S$ 
and $D_S$ and the octet mass in the chiral limit. Using the baryon masses and
the value of $\sigma_{\pi N} (0)$ as input, one can determine all LECs
in terms of one parameter, $\beta_R$. This parameter can be shown to be 
bounded and the  observables are insensitive to variations of it within 
its allowed range. Furthermore, it was also demonstrated
that the effects of two (and higher) loop diagrams can almost entirely
be absorbed in a redefinition of the one loop renormalization parameters. 
Within this scheme, one finds for the octet baryon mass in the chiral limit
$\krig{m} = 770\pm 110\, {\rm MeV}$. The quark mass expansion of the nucleon 
mass,  in the notation of Eq.(\ref{massform}), reads
\un{$m_N  = \,  \krig{m}  \, ( 1 + 0.34 - 0.35 + 0.24 \, )$}.
One observes that there are large cancellations between the second order
and the leading non--analytic terms of order $p^3$, a well--known effect.
From the chiral expansion of the nucleon mass one can not    
yet draw a final conclusion about the rate of convergence in the
three--flavor sector of baryon CHPT. 
The chiral expansion of the $\pi N$ $\sigma$--term shows a moderate 
convergence, i.e. the terms of increasing order become successively smaller,
$\sigma_{\pi N} (0) = 58.3 \, ( 1 - 0.56  + 0.33) \, \, \, {\rm MeV} 
= 45 \, \, {\rm MeV}$.
Still, the $p^4$ contribution is important.  For the strangeness fraction  
$y$ and $\hat \sigma$, one finds \un{$y = 0.21 \pm 0.20$, 
$\hat \sigma = 36 \pm 7 \, \, {\rm MeV}$}.
The value for $y$ is within the band deduced in~\cite{gls}, $y =
0.15 \pm 0.10$ and the value for $\hat \sigma$ compares favourably
with Gasser's estimate, $\hat \sigma = 33 \pm 5\,$MeV~\cite{jg}. Further 
results concerning the kaon--nucleon $\sigma$--terms and some two--loop
corrections to the nucleon mass can be found in~\cite{bm}.

\subsection*{The remainder at the Cheng--Dashen point and the scalar
  form factor}

\noindent Finally, two more
comments concerning the difference of the pion--nucleon $\sigma$--term 
at $t=0$ and at the Cheng--Dashen point are in order. First, in~\cite{bkmcd}
it was shown that the remainder $\Delta_R$ not fixed by chiral symmetry, 
i.e. the
difference between the on--shell $\pi N$ scattering amplitude $\bar{D}^+
(0,2M_\pi^2)$ and the scalar form factor $\sigma_{\pi N} (2M_\pi^2)$,
$\Delta_R \equiv F_\pi^2 \,\bar{D}^+(0,2M_\pi^2) -\sigma_{\pi N} (2M_\pi^2)$, 
contains no chiral logarithms and vanishes simply as $M_\pi^4$ in the
chiral limit. In addition, an upper limit was reported, 
\un{$\Delta_R \le 2\,$MeV}. This value is obtained as follows. The
chiral one--loop corrections give $\Delta_R^{\rm loop} =
0.5$~MeV~\cite{gss}, from the baryon resonances only the
$\Delta$--isobar contributes non--negligibly, $\Delta_R^{\Delta} 
= 0.6$~MeV~\cite{bpp}, and t--channel scalar meson exchange can be 
bounded based on dimensional arguments to give $\Delta_R^{S} \le 
1.1$~MeV~\cite{bkmcd}. In addition, from the $K\eta$ loop contribution
to the scalar form factor evaluated in~\cite{bkmmass}, one finds an
upper bound for the strangeness contribution,  $\Delta_R^{\rm strange}
\le 0.4$~MeV.  
Second, in~\cite{bkmmass} it was shown that a one--loop
diagram with an intermediate $\Delta (1232)$ allows to explain the numerical
value of the scalar form factor (ff). The leading $p^3$ graph with nucleon
intermediate states gives only $7.4\,$MeV, i.e. half of the empirical value
\cite{gls}. The $\Delta$-contribution, which formally starts at order $p^4$, adds
another $7.5\,$MeV. However, it was already stressed in~\cite{bkmmass} that
 the spectral function
Im~$\sigma_{\pi N} (t)/t^2$ is much less peaked around $t = 4M_\pi^2$
than the empirical one given in\cite{gls}. The 
$\Delta$-contribution enhances the tail of the spectral function  
at larger $t$,
in contrast to the strong pionic correlations (higher loop effects),
which tend to enhance the spectral function close to threshold. Furthermore,
the SU(3) calculation of ref.\cite{bm}  indicates fairly sizeable strangeness
effects in this quantity. More detailed higher order calculations are
necessary to clarify this issue.

\section*{SINGLE PION PRODUCTION}

\noindent In this section, I will make some comments on the status of
pion production by real and virtual photons as well as in $pp$
collisions.

\subsection*{Pion photoproduction off nucleons}
 
\noindent In table~2, the CHPT predictions for charged and neutral
pion production off nucleons are summarized in comparison to a
dispersion--theoretical analysis and the available data. The chiral
expansion converges quickly for the charged pions due to the
dominance of the Kroll-Ruderman term. For neutral pions, the situation
is more subtle. In fact, the threshold value of the electric dipole
amplitude for neutral pions off protons is not a good testing ground
of CHPT but rather the energy dependence in the threshold region. For
the case of the neutron, the situation is more favorable since the
first correction to the leading term $\sim M_\pi^2$ is only of the
order of  30\%.

\renewcommand{\arraystretch}{1.2}
\begin{table}[hbt]
\caption{Predictions and data for the electric dipole amplitudes.}
\vspace{0.2cm}
\label{tab:effluents}
\begin{tabular}{|l|c|c|c|}
\hline
                                & CHPT\protect{\cite{bkmcp}}
                                & DR\protect{\cite{dht}}     & Experiment  \\
\hline
$E_{0+}^{\rm thr} (\pi^+ n)$    & $28.2 \pm 0.6$ & $28.0$ 
                                & $27.9\pm 0.5$\protect{\cite{burg}},
                                $28.8 \pm 0.7$\protect{\cite{adam}} 
                                $27.6 \pm 0.3$\protect{\cite{salp}}  \\
$E_{0+}^{\rm thr} (\pi^- p)$    & $-32.7 \pm 0.6$ & $-31.7$ & $-31.4
                                \pm 1.3$\protect{\cite{burg}}, 
                               $-32.2 \pm 1.2$\protect{\cite{gold}}, 
                               $-31.5\pm 0.8$\protect{\cite{triumf}}  \\
$E_{0+}^{\rm thr} (\pi^0 p)$    & $-1.16$ & $-1.22$ 
                                & $-1.31\pm 0.08$\protect{\cite{fuchs}},
                                $-1.32\pm 0.11$\protect{\cite{berg}} \\
$E_{0+}^{\rm thr} (\pi^0 n)$    & $2.13$ & $1.19$ & 
                                $1.9 \pm 0.3$\protect{\cite{argan}} \\ 
\hline
\end{tabular}
\end{table}

\noindent The question arises how to measure the neutron amplitude?
The natural neutron target is the deuteron. The transition matrix for
$\pi^0$ production off the deuteron (d) takes the form
${\cal T} = 2i \, E_d \, {\vec J} \cdot \vec{\epsilon} + {\cal O}(\vec{q
  \,})$,
with ${\vec J}$ the total angular momentum of the d and
$\vec{\epsilon}$ the polarization vector of the photon. Although the
deuteron electric dipole amplitude could be calculated entirely within
CHPT, a more precise calculation is based on the approach suggested by
Weinberg~\cite{weinpid}, i.e. to calculate matrix elements of the type
$\langle \Psi_d | {\cal K} | \Psi_d\rangle$ by using deuteron wave
functions  $\Psi_d$ obtained from accurate  phenomenological NN potentials and
to chirally expand the kernel ${\cal K}$. Diagrammatically, one has
the single scattering (ss) terms which contain the desired $\pi^0 n$
amplitude. In addition, there are the so--called three--body (th)
contributions (meson exchange currents). To leading order $p^3$, one only
has the photon coupling to the pion in flight and the seagull
term~\cite{blvk}. The latter involves the charge exchange amplitude and is thus
expected to dominate the single scattering contribution. However, to
obtain the  same accuracy as for the ss terms, one has to calculate
also the corrections at order $p^4$. This has been done in
\cite{bblmvk}. It was shown that the next--to--leading order
three--body corrections and the possible four--fermion contact terms
do not induce any new unknown LEC and one therefore can calculate
$E_d$ in parameter--free manner. One finds
\beq \label{Ed}
E_d = E_d^{\rm ss} + E_d^{\rm tb,3} + E_d^{\rm tb,4} = 0.36 -1.90
-0.25 = (-1.8 \pm 0.2) \cdot 10^{-3}/M_{\pi^+} \,\,\, .
\eeq
Some remarks concerning this result are in order. First, one finds
indeed that the tb contribution is bigger than the single scattering
one. However, the former can be calculated precisely, i.e. the first
corrections amount to a meager 13\%. This signals good convergence.
I remark that a recent claim about large higher order (unitarity)
corrections \cite{pw} needs to be quantified in a consistent CHPT calculation.
Second, the resulting $E_d$ is very sensitive to $E_{0+} (\pi^0 n)$.
If one were to set $E_{0+} (\pi^0 n)= 0$, $E_d$ changes to $-2.6 
\cdot 10^{-3}/M_{\pi^+}$, i.e. the threshold cross sections would
change by a factor of two. Note that the theoretical error given in
Eq.(\ref{Ed}) is an educated guess, see \cite{bblmvk}.
Third, the CHPT prediction nicely agrees
with the empirical value of $E_d^{\rm exp} = (-1.7 \pm 0.2)\cdot
10^{-3}/M_{\pi^+}$~\cite{argan}. This agreement might, however, be 
fortitious since
the extraction of the empirical number relies on the input from the
elementary proton amplitude   to fix a normalization constant.  The
TAPS collaboration intends to redo this measurement at MAMI and
SAL is going to report their result very soon. The
consequences of precisely determine these S--wave amplitudes to
test isospin symmetry and its violation are discussed in~\cite{ulfos}.

\smallskip

\noindent A short comment on the P--waves in $\pi^0$ production is
in order. Here, the chiral series converges quickly and novel
\un{low--energy theorems (LETs)} for the multipole combinations $P_1 =
3E_{1+} + M_{1-} - M_{1+}$ and $P_2 =3E_{1+} - M_{1+} + M_{1-}$ have 
been derived in~\cite{bkmzpc}. While $P_1$ can be inferred from the
unpolarized data, extracting $P_2$ calls for polarization. Such an 
experiment has been done at MAMI but the analysis is not yet finished.
The LET for $P_1$ agrees with a remarkable accuracy with the data,
\beq
P_1^{\rm LET} = 10.33 \,\,\, , \quad P_1^{\rm exp} = 10.02 \pm 0.15 
\,\, \cite{fuchs} \,\,\, ,
\eeq
in conventional units. This
means that the P--waves are an \un{excellent testing ground} of the
chiral QCD dynamics, contrary to common folklore. The order $p^4$
corrections to these P--wave LETs are presently under 
investigation~\cite{bkmlp4}.
Finally, I note that Al Nathan and Hans Str\"oher independently
pointed out to me that the chiral predictions for $P_1$ and the
charged electric dipole amplitudes $E_{0+}(\pi^+n,\pi^-p)$ 
based on the latest determinations can
be brought into perfect agreement with the CHPT predictions
if one decreases the value of the 
pion--nucleon coupling constant to $g_{\pi N} = 13.06$. 
This deserves further study.

\subsection*{Neutral pion electroproduction off the proton}
 
\noindent Producing the pion with virtual photons offers further insight
since one can extract the longitudinal S--wave multipole $L_{0+}$ and also
novel P--wave multipoles. Data have been taken at
NIKHEF~\cite{welch}\cite{benno} 
and MAMI~\cite{distler} for
photon virtuality of $k^2 = -0.1$~GeV$^2$. In fact, it has been argued
previously that such photon four--momenta are already too large for
CHPT tests since the loop corrections are large~\cite{bklm}. However,
these calculations were performed in relativistic baryon CHPT and thus
it was necessary to redo them in the heavy fermion formalism. This was
done in~\cite{bkmel}. The abovementioned data for differential cross
sections were used to determine the three novel S--wave LECs. I should
mention that one of the operators used is of dimension five, i.e. one
order higher than the calculation was done. This can not be
circumvented since it was shown that the two S--waves are
overconstrained by a LET valid up to order $p^4$. The resulting
S--wave cross section $a_0 = |E_{0+}|^2 + \ve_L \,|L_{0+}|^2$ shown in  
fig.~1 is in fiar agreement with the data. Note also that it is
dominated completely by the $L_{0+}$ multipole (upper dot-dashed line)
since $E_{0+}$ passes through zero at $k^2 \simeq
-0.04$~GeV$^2$. However, in agreement with the older (and less precise)
calculations, the one loop corrections are large so one should compare at
lower photon virtualities. In ref.\cite{bkmel}, many predictions for
$k^2 \simeq -0.05$~GeV$^2$ are given, see fig.~2.
At MAMI, data have been taken in
this range of $k^2$ and we are looking forward to their analysis, in
particular it will be interesting to nail down the zero--crossing of
the electric dipole amplitude and to test the novel P--wave LETs~\cite{bkmprl}.

\parbox{8cm}{
\begin{center}
\epsfig{figure=figa0.epsi,width=7cm,height=6cm}
\end{center}}
\parbox{5.2cm}{\vspace*{2.2cm}
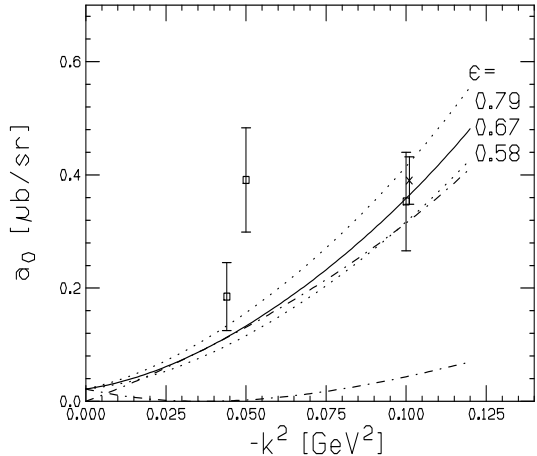
{\small \setlength{\baselineskip}{2.6ex} Fig.~1. The S--wave cross
  section $a_0$ for $\ve = 0.67$ (solid line) in comparison to the
  data. The dotted lines correspond to $\ve = 0.52$ and $0.79$,
  respectively. The upper dot--dashed is the contribution from 
  $\ve  \,|L_{0+}|^2$. Boxes:
  \protect{\cite{welch}}, cross: \protect{\cite{benno}}.}}

\parbox{8cm}{
\begin{center}
\epsfig{figure=figsec.epsi,width=7cm,height=8cm}
\end{center}}
\parbox{5.2cm}{\vspace*{3.2cm}
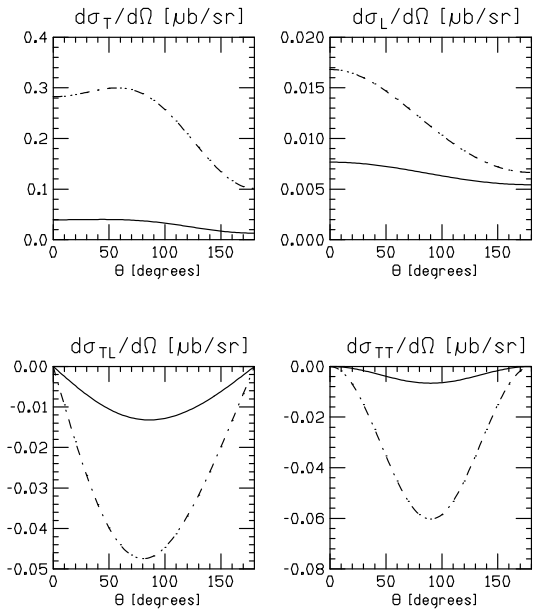
{\small \setlength{\baselineskip}{2.6ex} Fig.~2. Predictions for
 the various differential cross sections for $\gamma^\star p \to \pi^0 p$
 with  $\ve = 0.58$, $k^2 = -0.06$~GeV$^2$
 and $\Delta W = 2$ and $8$~MeV as indicated by the solid and
 dashed--dotted lines, respectively. Here, $\Delta W = W - W_{\rm
   thr}$ denotes the
 total cms energy above threshold, $W_{\rm thr} = m_p + M_{\pi^0} =
 1074.25$~MeV.}}

\subsection*{Pion production in proton--proton collisions}

\noindent The high precision data for the processes $pp \to  
pp\pi^0$ and $pp\to d\pi^+$ in the threshold
region~\cite{meyer}\cite{who} have spurred a flurry of theoretical
investigations. In particular, tree--level chiral perturbation theory
including dimension two operators has been used to constrain the
long--range pion--exchange contributions (the so--called direct, i.e.
the production off one nucleon, and
rescattering, i.e. the one--pion--exchange current, graphs) 
\cite{pmmmk,bira1,bira2,lee}. On the other hand,
this process has also been considered in the  successful
semi--phenomenological meson--exchange models, a particular example
being the one considered in ref.\cite{unsers}. In fact, the chiral
perturbation theory approach should lead to a deeper
understanding of the success of the meson--exchange picture, as first
stressed by Weinberg~\cite{wein}. It is therefore striking that the
calculations performed so far lead to a marked difference in the
role of the so--called recattering contribution, which interferes
constructively with the direct production in the J\"ulich model 
(this effect was already observed in \cite{oset})  and 
destructively in the chiral framework, respectively. Note that while there is
still debate about the actual numerical treatment (co-ordinate versus
momentum space) and the ensuing size of the rescattering contribution
in the CHPT approaches~\cite{lee}, the sign
difference to the meson--exchange model can be considered a genuine
feature. It is exactly this point which is addressed in~\cite{hhhms}.
I argue that the treatment underlying the isoscalar
pion--nucleon scattering amplitude and the related transition operator
for the process $NN \to NN\pi$ in the chiral framework is not yet 
sufficiently accurate and thus the resulting rescattering contribution
might be considered an artefact of this approximation. Clearly, this 
does not mean that CHPT is invalid but rather
that higher order  (one loop) effects need to be accounted for. To
make this statement most transparent, I will focus the attention
on the isoscalar $\pi N$ scattering amplitude. At second order in small
momenta, 
it takes the form (up to normalization factors, collectively 
denoted by ${\cal N}$)
\beq\label{tplus} 
 T^+ (q,k) = {\cal N} \,
\frac{M_\pi^2}{F_\pi^2} \biggl( 2c_1 - (c_2- \frac{g_A^2}{8m})
\frac{\omega_q \, \omega_k}{M_\pi^2} -c_3 \, \frac{q \cdot k}{M_\pi^2}  
 \biggr) +  T^+_{\rm Born} (q,k) + {\cal O}(M_\pi^3)
\,\,\, ,
\eeq
where $q \,(k) = (\omega_{q \, (k)} , \vec q \, (\vec k \,) )$ denotes
the four--momentum of the produced (exchanged) pion. 
The LECs $c_i$ have been determined to second~\cite{bkmppn} 
and third order~\cite{bkma}, respectively, 
\beqa\label{citree}
{\cal O}(p^2) &:& c_1 =  -0.64 \pm 0.14 \, \, , \quad
c_2  =  1.78 \pm 0.10 \, \, , \quad 
c_3  = -3.90 \pm 0.09 \, \, , \\ \label{ciloop}
{\cal O}(p^3) &:&
c_1 =  -0.93 \pm 0.10 \, \, , \quad
c_2  =  3.34 \pm 0.20 \, \, , \quad 
c_3  = -5.29 \pm 0.25 \, \, .
\eeqa
with all numbers given in GeV$^{-1}$. It is important to stress at order $p^2$,
i.e. to the accuracy used in describing $pp\to pp\pi^0$, it is \un{not} possible
to simultaneously fit the pertinent sub-- and threshold parameters. This is
only possible at one loop.

\parbox{8cm}{
\begin{center}
\epsfig{figure=np.epsi,width=7cm,height=6cm}
\end{center}}
\parbox{5.2cm}{\vspace*{1.8cm}
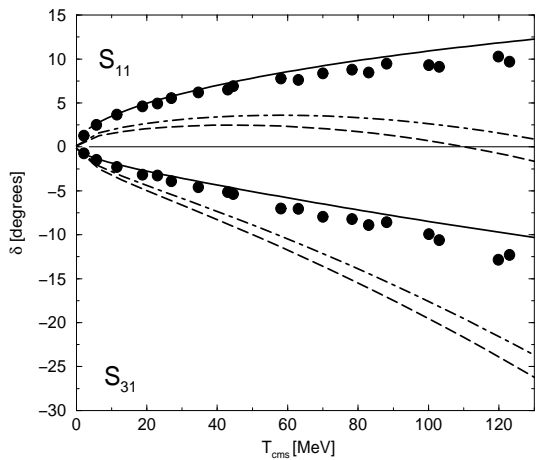
{\small \setlength{\baselineskip}{2.6ex} Fig.~3. The $\pi$N  phase shifts
  $S_{11}$ and $S_{31}$ based on the dimension two CHPT calculation
  using the parameters sets given in
  Eqs.\protect{\ref{citree},\ref{ciloop}} 
  (dashed and dashed--dotted lines) in comparison to the data and the fit
  within the J\"ulich meson--exchange model (solid lines).}}

\noindent In fig.~3, the corresponding $\pi$N phase shifts
$S_{11}$ and $S_{31}$ are shown in comparison to the data and the fit within 
the J\"ulich model.  The CHPT description based on Eq.\ref{tplus} and
using the ranges for the LECs as given in Eqs.\ref{citree},\ref{ciloop} 
can not be expected
to be sufficiently accurate for the threshold kinematics for pion 
production in $pp$ collisions, where $\vec{q} \simeq 0$ and 
$\vec k \simeq 370$~MeV. Indeed, when constructing the transition operator
for the rescattering contribution, this inadequacy carries over. 
Even if
one artificially changes the value of $c_2$ and/or $c_3$ to describe the
on--shell scattering data up to $T_{\rm cms} = 150$~MeV, the corresponding
rescattering graph still interferes destructively with the direct one.
In contrary, using the meson--exchange model,
which fits the $\pi N$ data up to energies of about 500~MeV,
this interference is constructive.
It is therefore conceivable  that the CHPT calculations
based on the dimension two, tree level approach to the isoscalar
scattering amplitude are not sufficiently accurate and that one has to
go to one loop before drawing any conclusion. However, there is a
\un{loophole} in this argument. In the meson--exchange model, crossing
symmetry is violated. The precise consequences of this need to be
investigated. Interestingly, the
reaction $pp \to d\pi^+$ is dominated by the isovector amplitude and
receives a much smaller contribution from heavy meson
exchange processes. This means that long--range (chiral) physics
should play a more dominant role and this fact remains to be studied
in detail. In particular, the comparison with the data should unravel
the relative sign and strength of the isovector and the isoscalar $\pi$N
amplitude entering the pion production operator.

\section*{SUMMARY AND OUTLOOK}

\noindent In this talk, I could only give a short glimpse on the many facets of
baryon chiral perturbation theory. Another interesting topics like 
$\pi N \to \pi\pi N$, chiral symmetry constraints on the NN interaction
and the physics underlying the LECs of the effective chiral
pion--nucleon Lagrangian (resonance exchange saturation) are treated
in Norbert Kaiser's talk~\cite{nk}. There has also been made
considerable progress in the three flavor sector, topics including 
renormalization~\cite{guido}, the chiral expansion of the octet
magnetic moments~\cite{sven2} or threshold kaon production off
protons~\cite{sven1}. In addition, a consistent scheme to implement
the $\Delta$--isobar in the effective field theory has recently
been proposed~\cite{hhk}. This is based on counting the
$\Delta$--nucleon mass difference, $m_\Delta - m_N \simeq 3F_\pi$,
as an additional small parameter. How good this phenomenologically
inspired approach really is remains to be seen. Also, much more work
than described here has been done in connection with nuclei.   
For a review about CHPT in few--nucleon systems, see van Kolck~\cite{bira3}.
Obviously, we need \un{more precise low--energy data}
to further test the chiral dynamics of QCD as exemplified in the
discussion about isospin violation in the pion--nucleon interaction
and the role of electromagnetic corrections.

\bibliographystyle{unsrt}

\begin{thebibliography}{99}
\bibitem{heiri} H. Leutwyler, Ann. Phys. (NY) {\bf 235}, 165 (1994).
\bibitem{bkmr} V. Bernard, N. Kaiser and Ulf-G. Mei{\ss}ner, 
 Int. J. Mod. Phys. {\bf E4}, 193 (1995). 
\bibitem{juerg} J. Gasser, in Proceedings ``Physics with Light
  Mesons and Second International Workshop on $\pi$N Physics", 
  Los Alamos Conference LA-11184-C, W.R. Gibbs and
  B.M.K. Nefkens (eds.),  1987.
\bibitem{bkma} V. Bernard, N. Kaiser and Ulf-G. Mei{\ss}ner,
  Nucl. Phys. {\bf A615}, 483 (1997). 
\bibitem{bkmpin} V. Bernard, N. Kaiser and Ulf-G. Mei{\ss}ner, 
Phys. Lett. {\bf B309}, 421 (1993); Phys. Rev. {\bf C52}, 2185 (1995).
\bibitem{sigg} D. Sigg et al.,  Nucl. Phys. {\bf A609}, 
269 (1996); (E) {\bf A617}, 526 (1997). 
\bibitem{bible} G. H\"ohler,  in  Landolt-B\"ornstein, Vol. 9b2, 
ed H. Schopper (Springer, Berlin, 1983).  
\bibitem{moj} M. Moj\v zi\v s, [hep-ph/9704415], Z. Phys. {\bf C} (1997), in
  print.
\bibitem{jg} J. Gasser, Ann. Phys.(NY) {\bf 136}, 62 
 (1981). 
\bibitem{bm} B. Borasoy and Ulf-G. Mei{\ss}ner,  Ann. Phys. (NY) {\bf
    254}, 192 (1997).
\bibitem{gls} J. Gasser, H. Leutwyler and M.E. Sainio,
Phys. Lett. {\bf B253}, 252, 260 (1991). 
\bibitem{bkmcd}V. Bernard, N. Kaiser and Ulf-G. Mei{\ss}ner, 
Phys. Lett. {\bf B389}, 144 (1996).
\bibitem{gss} J. Gasser, M.E. Sainio and A. ${\rm {\check S}}$varc,
Nucl. Phys.  {\bf B307}, 779 (1988).
\bibitem{bpp}L.S. Brown, W.J. Pardee and R.D. Peccei, Phys. Rev.
{\bf D4}, 2801 (1971).
\bibitem{bkmmass} V. Bernard, N. Kaiser and Ulf-G. Mei\ss ner, 
Z. Phys. {\bf C60}, 111 (1993). 
\bibitem{bkmzpc} V. Bernard, N. Kaiser and Ulf-G. Mei{\ss}ner,
Z. Phys. {\bf C70}, 483  (1996).
\bibitem{bkmcp} V. Bernard, N. Kaiser and Ulf-G. Mei{\ss}ner,
Phys. Lett. {\bf B393}, 116 (1996).
\bibitem{dht}O. Hanstein, D. Drechsel and L. Tiator, Phys. Lett. 
{\bf B399}, 13 (1997).
\bibitem{burg}J.P. Burg, Ann. Phys. (Paris) {\bf 10}, 363 (1965).
\bibitem{adam}M.J. Adamovitch et al., Sov. J. Nucl. Phys. {\bf 2}, 95 
 (1966).
\bibitem{salp}J. Bergstrom, private communication.
\bibitem{gold}E.L. Goldwasser et al., Proc. XII Int. Conf. on
  High--Energy Physics, Dubna, 1964, ed. Y.--A. Smorodinsky
  (Atomzidat, Moscow, 1966).
\bibitem{triumf}M.A. Kovash et al., $\pi$N Newsletter {\bf 12}, 51 (1997).
\bibitem{ulf95}Ulf-G. Mei{\ss}ner in "Baryons '95", B.F. Gibson et al.
(eds), World Scientific, Singapore, 1996.
\bibitem{bkme0p} V. Bernard, N. Kaiser and Ulf-G. Mei{\ss}ner,
Phys. Lett. {\bf B378}, 337  (1996).
\bibitem{fuchs} M. Fuchs et al., Phys. Lett. {\bf B368}, 20 (1996).
\bibitem{berg} J. Bergstrom et al., Phys. Rev. {\bf C53}, R1052  (1996). 
\bibitem{argan} P. Argan et al., Phys. Lett. {\bf B206}, 4 (1988).
\bibitem{weinpid}S. Weinberg, Phys.  Lett. {\bf B295}, 114 (1992). 
\bibitem{blvk}S.R. Beane, C.Y. Lee and U. van Kolck, Phys. Rev. {\bf
 C52}, 2914  (1995).
\bibitem{bblmvk}S.R. Beane, V. Bernard, T.-S.H. Lee,
  Ulf-G. Mei{\ss}ner and U. van Kolck, Nucl. Phys. {\bf A618}, 381 (1997).
\bibitem{pw} P. Wilhelm, Mainz preprint MKPH-97-8, 1997
  [nucl-th/9703037].
\bibitem{ulfos} Ulf-G. Mei{\ss}ner, J\"ulich preprint KFA-IKP(TH)-1997-08, 
[hep-ph/9706367]. 
\bibitem{bklm}V. Bernard, N. Kaiser, T.-S.H. Lee  and
  Ulf-G. Mei{\ss}ner, Phys. Rep. {\bf 246}, 315 (1994).
\bibitem{bkmel}V. Bernard, N. Kaiser and Ulf-G. Mei{\ss}ner,
  Nucl. Phys. {\bf A607}, 379 (1996).
\bibitem{bkmlp4} V. Bernard, N. Kaiser and Ulf-G. Mei{\ss}ner, in preparation.
\bibitem{welch}T.P. Welch et al., Phys. Rev. Lett. {\bf 69}, 2761 (1992).
\bibitem{benno}H.B. van den Brink et al., Phys. Rev. Lett. {\bf 74},
  3561 (1995); Nucl. Phys. {\bf A612}, 391 (1997).
\bibitem{distler}M. Distler, Thesis, University of Mainz, 1996; 
Th, Walcher, in "Baryons '95", B.F. Gibson et al.
(eds), World Scientific, Singapore, 1996.
\bibitem{bkmprl}V. Bernard, N. Kaiser and Ulf-G. Mei{\ss}ner,
Phys. Rev. Lett. {\bf 74}, 3752 (1995).
\bibitem{meyer} H.O. Meyer et al. Phys. Rev. Lett. {\bf 65}, 2846 (1990);
Nucl. Phys. {\bf A539}, 663 (1992).
\bibitem{who} M. Drochner et al., Phys. Rev. Lett. {\bf 77}, 454
  (1996); C. Heimberg et al., Phys. Rev. Lett. {\bf 77}, 1012 (1996). 
\bibitem{pmmmk} B.Y. Park, F. Myhrer, J.R. Morones, T. Meissner and
K. Kubodera,  Phys. Rev. {\bf C53}, 1519 (1996).
\bibitem{bira1} T.D. Cohen, J.L. Friar, G.A. Miller and U. van Kolck,
 Phys. Rev. {\bf C53}, 2661 (1996).
\bibitem{bira2}  U. van Kolck, G.A. Miller and D.O. Riska,
Phys. Lett. {\bf B388}, 679 (1996).
\bibitem{lee} T. Sato, T.-S.H. Lee, F. Myhrer and K. Kubodera,
[nucl-th/9704003].
\bibitem{unsers}
C. Hanhart, J. Haidenbauer, A. Reuber, C. Sch\"utz and J. Speth,
Phys. Lett. {\bf B358} 21 (1995).
\bibitem{wein} S. Weinberg, Phys. Lett. {\bf B251}, 288 (1990).
\bibitem{oset} E. Hernandez and E. Oset, Phys. Lett. {\bf B350}, 158 (1995).
\bibitem{hhhms}  C. Hanhart, J. Haidenbauer, M. Hoffmann, Ulf-G. Mei{\ss}ner
and J. Speth, preprint KFA-IKP(TH)-1997-14 [nucl-th/9707029].
\bibitem{bkmppn} V. Bernard, N. Kaiser and Ulf-G. Mei{\ss}ner,
Nucl. Phys. {\bf B457}, 147 (1995).
\bibitem{nk} Norbert Kaiser, these proceedings.
\bibitem{guido} G. M\"uller and Ulf-G. Mei{\ss}ner,
Nucl. Phys. {\bf B492}, 379 (1997).
\bibitem{sven2}Ulf-G. Mei{\ss}ner and S. Steininger,  
Nucl. Phys. {\bf B499}, 349 (1997).
\bibitem{sven1}S. Steininger  and Ulf-G. Mei{\ss}ner, Phys. Lett.
{\bf  B391}, 461 (1997).
\bibitem{hhk} T.R.~Hemmert, B.R.~Holstein and J.~Kambor,
  Phys. Lett. {\bf B395}, 89 (1997).
\bibitem{bira3} U. van Kolck, Seattle preprint [hep-ph/9707228].

\end{thebibliography}

\end{document}